%% file: main.tex
\documentclass[manuscript]{acmart}

\usepackage{caption}
\usepackage{subcaption}
\usepackage{wrapfig}
\usepackage{pdfpages}
\usepackage{algorithm}
\usepackage{algpseudocode}
\usepackage{longtable}
\usepackage{caption}
\usepackage{subcaption}
\usepackage{booktabs}
\usepackage{amsmath}
\usepackage{xcolor}
\usepackage{hyperref}
\usepackage{multirow} 

\usepackage{makecell}

\usepackage{ulem}
\newcommand\crossout{\bgroup\markoverwith{\textcolor{cyan}{\rule[0.5ex]{1pt}{1.5pt}}}\ULon}

\AtBeginDocument{%
  }

\begin{document}

\title{Detecting the Use of Generative AI in Crowdsourced Surveys: Implications for Data Integrity}

\begin{abstract}
\input{sections/abstract}
\end{abstract}




\author{Dapeng Zhang}
\affiliation{%
  \institution{The University of Tulsa}
  \city{Tulsa}
  \country{USA}}
\email{daz4358@utulsa.edu}

\author{Marina Katoh}
\affiliation{%
  \institution{The University of Tulsa}
  \city{Tulsa}
  \country{USA}
}
\email{mak5308@utulsa.edu}

\author{Weiping Pei}
\affiliation{%
 \institution{The University of Tulsa}
 \city{Tulsa}
 \country{USA}}
\email{weiping-pei@utulsa.edu}

\maketitle

\section{Introduction}
\input{sections/introduction}

\label{sec:introduction}


\section{Methodology}
\label{sec:methodology}
\input{sections/methodology}

\section{Results and Analysis}
\label{sec:results_analysis}
\input{sections/results}

\section{Conclusion}
\label{sec:conclusion}
\input{sections/conclusion}




\bibliographystyle{ACM-Reference-Format}
\bibliography{llm_crowdsourcing}

\end{document}

%% file: sections/abstract.tex
The widespread adoption of generative AI (GenAI) has introduced new challenges in crowdsourced data collection, particularly in survey-based research. While GenAI offers powerful capabilities, its unintended use in crowdsourcing, such as generating automated survey responses, threatens the integrity of empirical research and complicates efforts to understand public opinion and behavior. In this study, we investigate and evaluate two approaches for detecting AI-generated responses in online surveys: LLM-based detection and signature-based detection. We conducted experiments across seven survey studies, comparing responses collected before 2022 with those collected after the release of ChatGPT. Our findings reveal a significant increase in AI-generated responses in the post-2022 studies, highlighting how GenAI may silently distort crowdsourced data. This work raises broader concerns about evolving landscape of data integrity, where GenAI can compromise data quality, mislead researchers, and influence downstream findings in fields such as health, politics, and social behavior. By surfacing detection strategies and empirical evidence of GenAI's impact, we aim to contribute to ongoing conversation about safeguarding research integrity and supporting scholars navigating these methodological and ethical challenges. 

%% file: sections/introduction.tex
The release of ChatGPT~\cite{chatgpt2024} in 2022 is transforming various domains by democratizing access to generative AI (GenAI) and seamlessly integrating it into daily work and personal lives~\cite{wolf2024chatgpt, choi2024unlock}. However, the use of GenAI  raises concerns about data quality and integrity in crowdsourcing, particularly for subjective tasks that rely on individual experiences and perspectives. In this paper, we investigate the use of GenAI in crowdsourcing, focusing on online survey studies. Specifically, irresponsible workers may exploit GenAI to complete open-ended survey questions with minimal effort, compromising the integrity of crowdsourced data and resulting in misleading findings in human-subject studies. 
Although platforms like Prolific~\cite{prolific_llm} and some studies have recognized the use of GenAI in crowdsourcing~\cite{powers2024can}, effective detection mechanisms remain largely unexplored. To fill this gap, we investigate and evaluate two approaches for detecting AI-generated responses in online surveys: LLM-based detection and signature-based detection.  
Our experiments on seven survey studies reveal a significantly higher percentage of AI-generated responses in post-2022 studies compared to pre-2022 studies. Besides, signature-based detection effectively identifies not only AI-generated responses but also irrelevant responses. These findings highlight the widespread use of GenAI in crowdsourcing since the release of ChatGPT, emphasizing the urgent need for robust strategies to mitigate its impact on data quality and integrity in crowdsourcing.


%% file: sections/methodology.tex


\subsection{Overview of Design}
\begin{figure}
  \centering
  \scalebox{0.75}{
  \includegraphics[width=\textwidth]{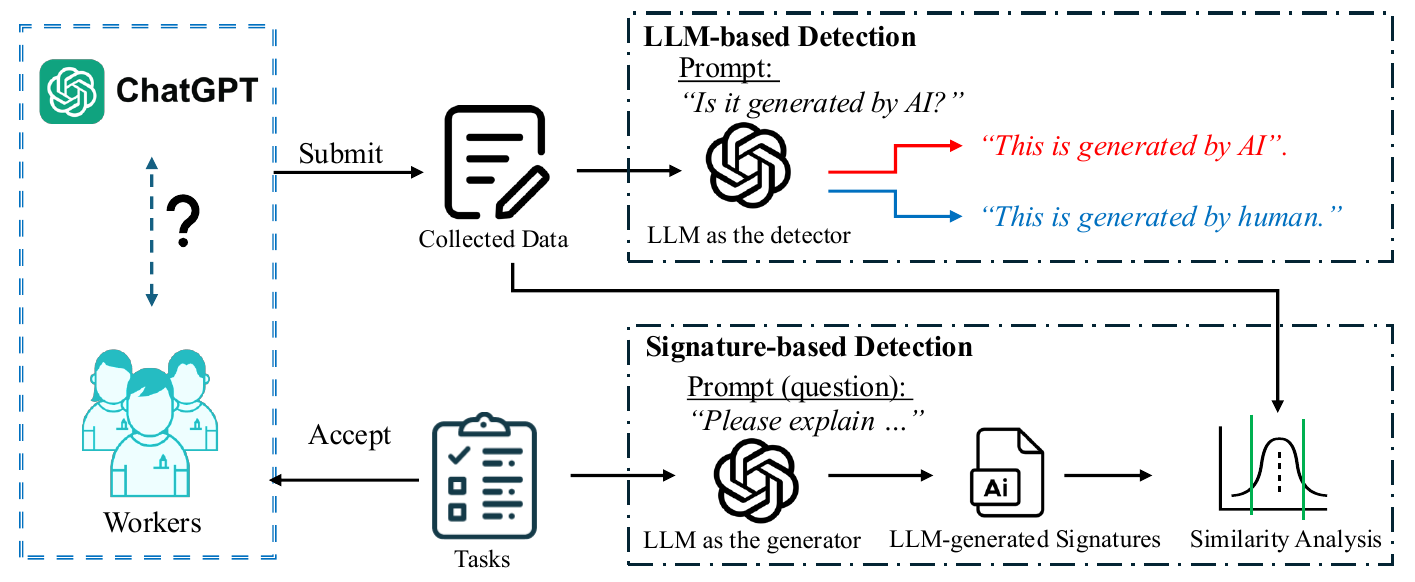}
  }
  \vspace{-0.2cm}
  \caption{Overview of Proposed Approaches for Detecting the Use of GenAI in Crowdsourcing.}
  \label{fig:overview}
  \vspace{-0.4cm}
\end{figure}

To detect the use of GenAI in crowdsourcing, we propose two complementary detection approaches: \textbf{\textit{LLM-based detection}} and \textbf{\textit{signature-based detection}}, as illustrated in Figure~\ref{fig:overview}. 
Our LLM-based detection approach is inspired by recent advances in LLM-generated text detection~\cite{zhu2023beat, koike2024outfox, tang2024science}. We adopt a zero-shot paradigm, in which we use \textbf{\textit{LLMs as detectors}}~\cite{mitchell2023detectgpt, zhu2023beat}. This approach has been extensively explored across various NLP tasks; however, to the best of our knowledge, its effectiveness in identifying AI-generated responses in online surveys remains underexplored. Specifically, we directly query LLMs to determine whether a given message is generated by AI~\cite{bhattacharjee2024fighting}. 
On the other hand, our signature-based detection is specifically designed for detecting AI-generated survey responses by utilizing \textbf{\textit{LLMs as generators}} rather than detectors. Drawing from an adversarial perspective, we leverage LLMs to generate potential AI-generated responses to survey questions before deploying the survey. These generated responses serve as reference signatures. We then conduct a similarity analysis to compare the collected responses with these signatures, based on the assumption that a higher similarity score indicates a greater likelihood of AI generation. 

\vspace{5pt}

\subsection{Experimental Setup}
\textbf{Survey data.} We focus on crowdsourced surveys, given their widespread use in both research and real-world applications, specifically evaluating responses to open-ended questions. To assess the effectiveness of our proposed detection approaches, we analyze seven survey studies as summarized in Table~\ref{table:datasets}. To ensure the generalizability of our findings, these studies span multiple domains and were conducted over a broad time frame, ranging from 03/21 to 06/24. Four of the studies were conducted in 2021, prior to ChatGPT's public release, allowing us to reasonably assume that their responses were generated without the influence of GenAI~\cite{natarajandetecting}. The remaining three studies were conducted between 03/23 and 06/24, a period during which GenAI tools were widely accessible. Additionally, five of the seven studies have been published in premier conferences or journals, and three were conducted by researchers outside the author team.

\begin{table}[h]
\caption{Summary of Survey Studies. Studies \#5 and \#6 are currently under review at premier journals.}
\label{table:datasets}
\scalebox{0.9}{
\begin{tabular}{c|c|c|c|c|c}
\hline
\begin{tabular}[c]{@{}c@{}}\textbf{Survey} \end{tabular} & \begin{tabular}[c]{@{}c@{}}\textbf{Month/}\\ \textbf{Year}\end{tabular} & \begin{tabular}[c]{@{}c@{}}\textbf{Number of} \\ \textbf{Responses}\end{tabular} & \textbf{Domain}        & \begin{tabular}[c]{@{}c@{}}\textbf{Avg. Length}\end{tabular} & \textbf{Platform}                                                     \\ \hline
\#1~\cite{alshehri2023exploring}                                                     & 03/21                                                 & 798                                                            & IoT      &   30.21  $\pm$ 11.63                                                           & \begin{tabular}[c]{@{}c@{}}MTurk\end{tabular} \\ \hline
\#2~\cite{pei2023tale}                                                     & 06/21                                                 & 170                                                            & Software (App User)      &   30.43  $\pm$ 20.70                                                            & \begin{tabular}[c]{@{}c@{}}Reddit/Craigslist\end{tabular} \\ \hline
\#3~\cite{pei2023tale}                                                     & 06/21                                                 & 313                                                            & Software (Non App User)      &  29.24  $\pm$ 20.12                                                              & \begin{tabular}[c]{@{}c@{}}Reddit/Craigslist\end{tabular} \\ \hline
\#4~\cite{pei2023tale}                                                     & 09/21                                                 & 160                                                            & Crowdsourcing & 33.34   $\pm$ 23.77                                                             & MTurk                                                        \\ \hline
\#5~\cite{traylor2025threat}                                                     & 07/23                                                 & 1166                                                           & Energy        &   13.94 $\pm$ 13.35                                                              & MTurk                                                        \\ \hline
\#6                                                     & 03/24                                                 & 198                                                            & Software      &  23.28 $\pm$ 14.54                                                              & Prolific                                                     \\ \hline
\#7~\cite{PEI2025103089}                                                     & 06/24                                                 & 520                                                            & Cybercrime    &  32.10 $\pm$ 17.91                                                               & MTurk                                                        \\ \hline
\end{tabular}
}
\vspace{-0.4cm}
\end{table}

\textbf{LLMs and similarity analysis.} To evaluate detection performance across different LLMs, we consider four models in our experiments for both detection approaches: GPT-3.5-Turbo, GPT-4, GPT-4o, and GPT-4o-Mini. For the signature-based approach, we employ two prompt strategies to generate AI-generated responses as signatures: the basic prompt, which directly queries LLMs with survey questions, and the sentiment-based prompt, which prompts LLMs with survey questions while specifying a sentiment constraint to generated various signatures. To compare collected responses with AI-generated signatures, we compute text embeddings using the Sentence Transformer (SBERT) model~\cite{reimers-2019-sentence-bert} and measure similarity scores using cosine similarity. For each collected response, we obtain multiple similarity scores based on the different signatures generated by various settings\footnote{We consider four LLMs with five different temperatures ($t=0, 0.25, 0.5, 0.75, 1.0$) to get 20 signatures for each question.} and consider the highest similarity score as the final metric to determine the likelihood that a collected response is AI-generated.   

%% file: sections/results.tex
\textbf{LLM-based detection.}
The percentage of detected AI-generated responses is shown in~Table~\ref{table:results_llm_detection}. 
Among the four models evaluated, GPT-3.5-Turbo exhibited the lowest detection rate of AI-generated responses in studies conducted prior to 2022. Given that LLMs were not widely accessible to the public before 2022, we assume that the detected AI-generated responses in these studies are false positives. Consequently, GPT-3.5-Turbo achieves the best performance, with an average false positive rate of 6.16\% across the four pre-2022 studies. In contrast, GPT-4 and GPT-4o produced false positive rates exceeding 40\%. These findings align with the results from~\cite{bhattacharjee2024fighting}, which reported false positive rates of approximately 4\% and 6\% for two NLP datasets using GPT-3.5 and concluded that GPT-3.5 outperforms GPT-4 in distinguishing AI-generated text from human-written text.  
Although ground truth labels are unavailable for studies conducted after 2022, we observe a significant increase in detected AI-generated responses compared to pre-2022 studies. On average, 30.55\% of responses in post-2022 studies were identified as AI-generated, which is consistent with recent findings by~\cite{veselovsky2023prevalence}, stating that ``around 30\% used LLMs.''  These results demonstrate the increasing prevalence of AI-generated responses in surveys since 2022, underscoring the urgent need for effective detection strategies to ensure the integrity of crowdsourced survey data. 
\begin{table}[h]
\centering
\vspace{-0.2cm}
\caption{Percentage of AI-generated Responses Identified Using Zero-shot LLM-based Detection}
\label{table:results_llm_detection}
\vspace{-0.2cm}
\scalebox{0.85}{
\begin{tabular}{|c|ccccc|cccc|}
\hline
\multirow{2}{*}{\textbf{LLM model}} & \multicolumn{5}{c|}{\textbf{Before 2022}}                                                                                                                                                & \multicolumn{4}{c|}{\textbf{After 2022}}                                                                                                        \\ \cline{2-10} 
& \multicolumn{1}{c|}{\textbf{\#1 (798)}} & \multicolumn{1}{c|}{\textbf{\#2 (170)}} & \multicolumn{1}{c|}{\textbf{\#3 (313)}} & \multicolumn{1}{c|}{\textbf{\#4 (160)}} & \textbf{Average} & \multicolumn{1}{c|}{\textbf{\#5 (1166)}} & \multicolumn{1}{c|}{\textbf{\#6 (198)}} & \multicolumn{1}{c|}{\textbf{\#7 (520)}} & \textbf{Average} \\ \hline
\textbf{GPT-3.5-Turbo}              & \multicolumn{1}{c|}{13.91\%}            & \multicolumn{1}{c|}{1.18\%}             & \multicolumn{1}{c|}{7.03\%}             & \multicolumn{1}{c|}{2.50\%}             & 6.16\%           & \multicolumn{1}{c|}{48.71\%}             & \multicolumn{1}{c|}{18.18\%}            & \multicolumn{1}{c|}{24.76\%}            & 30.55\%          \\ \hline
\textbf{GPT-4}                      & \multicolumn{1}{c|}{56.02\%}            & \multicolumn{1}{c|}{35.88\%}            & \multicolumn{1}{c|}{37.70\%}            & \multicolumn{1}{c|}{39.38\%}            & 42.25\%          & \multicolumn{1}{c|}{32.08\%}             & \multicolumn{1}{c|}{62.12\%}            & \multicolumn{1}{c|}{66.77\%}            & 53.66\%          \\ \hline
\textbf{GPT-4o}                     & \multicolumn{1}{c|}{58.52\%}            & \multicolumn{1}{c|}{32.94\%}            & \multicolumn{1}{c|}{37.70\%}            & \multicolumn{1}{c|}{35.62\%}            & 41.20\%          & \multicolumn{1}{c|}{51.74\%}             & \multicolumn{1}{c|}{49.49\%}            & \multicolumn{1}{c|}{66.59\%}            & 55.94\%          \\ \hline
\textbf{GPT-4o-Mini}                & \multicolumn{1}{c|}{22.81\%}            & \multicolumn{1}{c|}{5.88\%}             & \multicolumn{1}{c|}{4.79\%}             & \multicolumn{1}{c|}{6.25\%}             & 9.93\%           & \multicolumn{1}{c|}{58.58\%}             & \multicolumn{1}{c|}{28.79\%}            & \multicolumn{1}{c|}{28.46\%}            & 38.61\%          \\ \hline
\end{tabular}
}
\vspace{-0.2cm}
\end{table}

\textbf{Signature-based detection.} We applied the proposed signature-based detection approach to all studies except Study \#1, which focused on interaction-based responses and was therefore excluded from this analysis. Table~\ref{table:results_signature} presents the percentage of detected AI-generated responses across different similarity thresholds. 
Our results indicated that the proportion of detected AI-generated responses is significantly higher in post-2022 studies than in pre-2022 studies when the similarity threshold exceeds 0.8. This suggests that a larger proportion of collected responses in post-2022 studies resemble AI-generated signatures, i.e., pre-generated responses obtained from LLMs. Specifically, when the similarity threshold is set to 0.9, our signature-based detection identifies 0.16\% and 0.47\% of responses as AI-generated using the basic prompt and sentiment-based prompt, respectively, in pre-2022 studies. In contrast, the corresponding detection rates for post-2022 studies are 2.12\% and 3.29\%, respectively. Assuming that no responses in pre-2022 studies were AI-generated, these results suggest that the false positive rates of signature-based detection remains low for pre-2022 studies, with the basic prompt outperforming the sentiment-based prompt in minimizing false positives. 


\begin{center}
\begin{minipage}{\textwidth}
\begin{minipage}[c]{0.56\textwidth}
    \centering
    \captionof{table}{Percentage of AI-generated Responses Identified Using Signature-based Detection. (basic: basic prompt, senti: sentiment-based prompt, th: similarity threshold)}
    \vspace{-0.2cm}
    \label{table:results_signature}
    \scalebox{0.68}{
    \begin{tabular}{|c|c|cccc|cccc|}
    \hline
    \multirow{2}{*}{\textbf{\begin{tabular}[c]{@{}c@{}}th\end{tabular}}} & \multirow{2}{*}{\textbf{\begin{tabular}[c]{@{}c@{}}Prompt\\ Strategy\end{tabular}}} & \multicolumn{4}{c|}{\textbf{Before 2022}}                                                                                    & \multicolumn{4}{c|}{\textbf{After 2022}}                                                                                     \\ \cline{3-10} 
    &                                                                                     & \multicolumn{1}{c|}{\textbf{\#2}} & \multicolumn{1}{c|}{\textbf{\#3}} & \multicolumn{1}{c|}{\textbf{\#4}} & \textbf{Average} & \multicolumn{1}{c|}{\textbf{\#5}} & \multicolumn{1}{c|}{\textbf{\#6}} & \multicolumn{1}{c|}{\textbf{\#7}} & \textbf{Average} \\ \hline
    \multirow{2}{*}{\textbf{0.7}}                                                            & \textbf{basic}                                                                      & \multicolumn{1}{c|}{15.88\%}      & \multicolumn{1}{c|}{7.35\%}       & \multicolumn{1}{c|}{13.12\%}      & \textbf{11.04\%} & \multicolumn{1}{c|}{6.09\%}       & \multicolumn{1}{c|}{14.65\%}      & \multicolumn{1}{c|}{15.74\%}      & 9.66\%           \\ \cline{2-10} 
    & \textbf{senti}                                                                      & \multicolumn{1}{c|}{23.53\%}      & \multicolumn{1}{c|}{13.10\%}      & \multicolumn{1}{c|}{21.25\%}      & \textbf{17.88\%} & \multicolumn{1}{c|}{7.55\%}       & \multicolumn{1}{c|}{21.72\%}      & \multicolumn{1}{c|}{18.04\%}      & 11.94\%          \\ \hline
    \multirow{2}{*}{\textbf{0.75}}                                                           & \textbf{basic}                                                                      & \multicolumn{1}{c|}{8.82\%}       & \multicolumn{1}{c|}{4.47\%}       & \multicolumn{1}{c|}{6.25\%}       & 6.07\%           & \multicolumn{1}{c|}{3.00\%}       & \multicolumn{1}{c|}{7.58\%}       & \multicolumn{1}{c|}{14.59\%}      & \textbf{6.68\%}  \\ \cline{2-10} 
    & \textbf{senti}                                                                      & \multicolumn{1}{c|}{11.76\%}      & \multicolumn{1}{c|}{6.39\%}       & \multicolumn{1}{c|}{11.88\%}      & \textbf{9.18\%}  & \multicolumn{1}{c|}{3.43\%}       & \multicolumn{1}{c|}{15.15\%}      & \multicolumn{1}{c|}{15.74\%}      & 8.06\%           \\ \hline
    \multirow{2}{*}{\textbf{0.8}}                                                            & \textbf{basic}                                                                      & \multicolumn{1}{c|}{1.18\%}       & \multicolumn{1}{c|}{1.92\%}       & \multicolumn{1}{c|}{1.25\%}       & 1.56\%           & \multicolumn{1}{c|}{2.23\%}       & \multicolumn{1}{c|}{2.53\%}       & \multicolumn{1}{c|}{13.82\%}      & \textbf{5.46\%}  \\ \cline{2-10} 
    & \textbf{senti}                                                                      & \multicolumn{1}{c|}{3.53\%}       & \multicolumn{1}{c|}{4.47\%}       & \multicolumn{1}{c|}{4.38\%}       & 4.20\%           & \multicolumn{1}{c|}{2.23\%}       & \multicolumn{1}{c|}{7.58\%}       & \multicolumn{1}{c|}{14.78\%}      & \textbf{6.26\%}  \\ \hline
    \multirow{2}{*}{\textbf{0.85}}                                                           & \textbf{basic}                                                                      & \multicolumn{1}{c|}{0.00\%}       & \multicolumn{1}{c|}{0.64\%}       & \multicolumn{1}{c|}{0.00\%}       & 0.31\%           & \multicolumn{1}{c|}{1.89\%}       & \multicolumn{1}{c|}{0.00\%}       & \multicolumn{1}{c|}{11.32\%}      & \textbf{4.30\%}  \\ \cline{2-10} 
    & \textbf{senti}                                                                      & \multicolumn{1}{c|}{0.00\%}       & \multicolumn{1}{c|}{2.24\%}       & \multicolumn{1}{c|}{1.25\%}       & 1.40\%           & \multicolumn{1}{c|}{1.97\%}       & \multicolumn{1}{c|}{3.54\%}       & \multicolumn{1}{c|}{13.05\%}      & \textbf{5.20\%}  \\ \hline
    \multirow{2}{*}{\textbf{0.9}}                                                            & \textbf{basic}                                                                      & \multicolumn{1}{c|}{0.00\%}       & \multicolumn{1}{c|}{0.32\%}       & \multicolumn{1}{c|}{0.00\%}       & 0.16\%           & \multicolumn{1}{c|}{1.29\%}       & \multicolumn{1}{c|}{0.00\%}       & \multicolumn{1}{c|}{4.80\%}       & \textbf{2.12\%}  \\ \cline{2-10} 
    & \textbf{senti}                                                                      & \multicolumn{1}{c|}{0.00\%}       & \multicolumn{1}{c|}{0.96\%}       & \multicolumn{1}{c|}{0.00\%}       & 0.47\%           & \multicolumn{1}{c|}{1.37\%}       & \multicolumn{1}{c|}{0.51\%}       & \multicolumn{1}{c|}{8.64\%}       & \textbf{3.29\%}  \\ \hline
    \end{tabular}
    }
\end{minipage}
\hfill
\begin{minipage}[c]{0.4\textwidth}
    \centering
    \includegraphics[width=\textwidth]{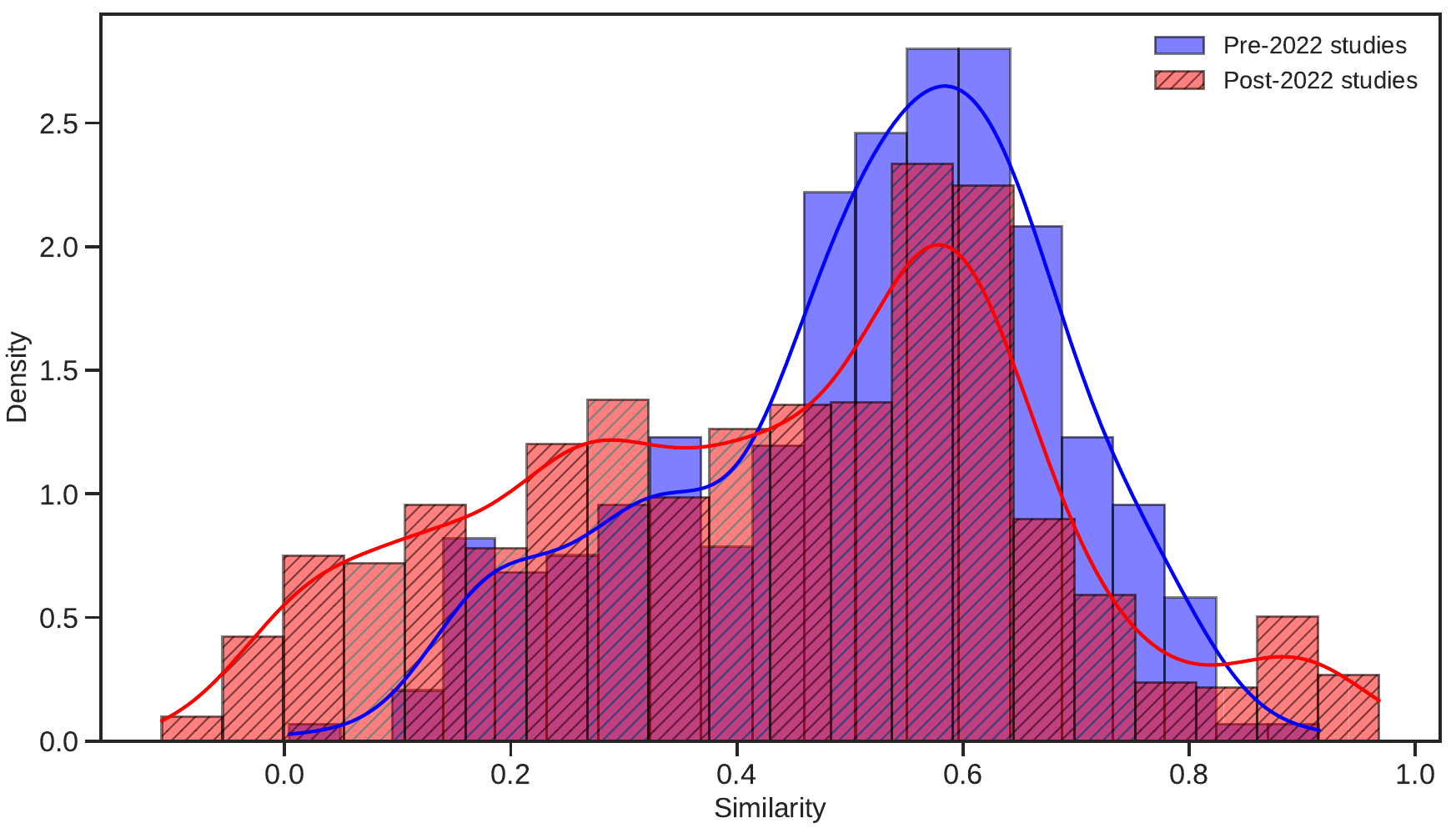}
    \vspace{-0.8cm}
    \captionof{figure}{Distribution of Similarity between Collected Responses and Signatures for Pre-2022 and Post-2022 Studies.}
    \label{fig:distribution_similarity}
\end{minipage}
\end{minipage}
\end{center}

\begin{figure}[H]
  \centering
  \scalebox{0.9}{
  \includegraphics[width=\textwidth]{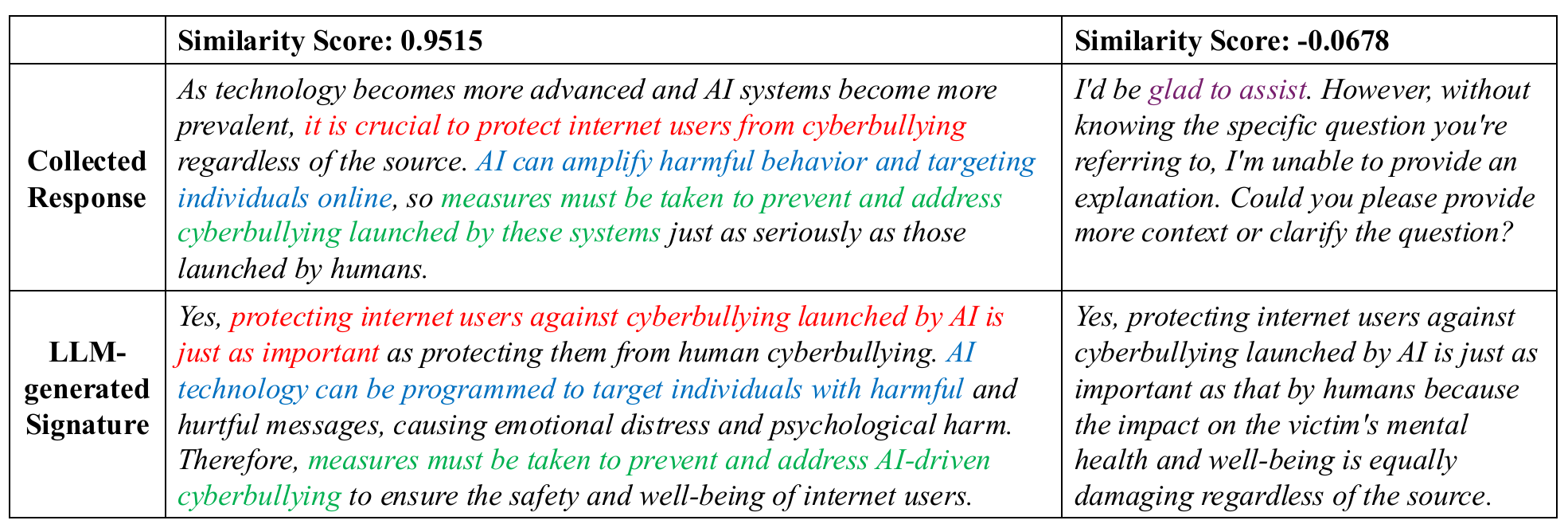}
  }
  \caption{Examples of AI-generated Responses and Signatures with High and Low Similarity Scores.}
  \label{fig:case_study}
\end{figure}

\textbf{Case study.} Figure~\ref{fig:distribution_similarity} presents the distribution of similarity scores for pre-2022 and post-2022 studies using the basic prompt strategy. As expected, post-2022 studies shows a higher proportion of responses with similarity scores close to 1. Upon manual inspection, we observed a substantial textual and semantic similarity between these responses and LLM-generated signatures. An illustrative example is provided in Figure~\ref{fig:case_study}. Interestingly, we also observed a higher proportion of responses with similarity scores close to 0 or even negative values in the post-2022 studies. Further manual analysis indicated that these responses were often irrelevant to the survey questions. One such example, shown in Figure~\ref{fig:case_study}, includes the phrase ``glad to assist'' and exhibits a conversational style, which suggest the use of GenAI. 
These findings indicate that signature-based detection not only effectively identifies AI-generated responses but also serves as a useful tool for spotting irrelevant or off-topic responses, further enhancing data quality in crowdsourcing. 

%% file: sections/conclusion.tex
In this work, we examine the growing presence of GenAI in crowdsourcing and evaluate two detection strategies: LLM-based and signature-based approaches. Through analysis of data from seven online surveys with open-ended questions, we observe a marked increase in AI-generated responses in the post-2022 studies compared to pre-2022 studies. Our findings highlight the evolving risks that GenAI pose to data integrity in participatory research contexts. Notably, signature-based detection not only identifies AI-generated content but also flags low-quality or irrelevant responses, offering practical value for maintaining research quality. These results underscore the urgent need for the CSCW community to develop robust, adaptive methods for safeguarding empirical research against emerging forms of automated content injection.